\documentclass{article}

\usepackage{amsmath}
\usepackage{amssymb}
\usepackage{amsxtra,color}
\usepackage{latexsym}
\usepackage{graphicx}
\usepackage{natbib}

\newcommand{\ket}[1]{| \: #1 \: \rangle}

\newcommand{\braOket}[3]{\langle \: #1 \: | \: #2 \:| \: #3 \: \rangle}

\begin{document}

\title{Diagrammatic approach to \\ attosecond delays in photoionization}

\author{J.~M.~Dahlstr\"om, T.~Carette and E.~Lindroth \\
\small Department of Physics, Stockholm University, \\ 
\small AlbaNova University Center, SE-106 91 Stockholm, Sweden}

\date{}

\maketitle 


\begin{abstract}
We study laser-assisted photoionization by attosecond pulses using a time-independent formalism based on 
diagrammatic many-body perturbation theory.
Our aim is to provide an {\it ab inito} route to the ``delays'' 
for this above-threshold ionization process,
which is essential for a quantitative understanding of attosecond metrology. 
We present 
correction curves for characterization schemes of attosecond pulses, such as ``streaking'', 
that account for the delayed atomic response in ionization from neon and argon. 
We also verify that photoelectron delays from many-electron atoms can be measured using similar schemes 
if, instead, the so-called continuum--continuum delay is subtracted. 
Our method is general and it can be extended also to more complex systems and additional correlation effects can be introduced systematically. 
\end{abstract}

\newpage

\section*{Introduction}
The temporal aspects of photoionization from many-electron systems can now be explored using extreme ultraviolet (XUV) attosecond pump pulses and infrared (IR) femtosecond laser probe fields.  
Experimental studies have evidenced relative delays between different ionization processes of various target systems 
including both solid-state surfaces 
\cite{CavalieriNature2007} 
and atomic gases 
\cite{SchultzeScience2010,KlunderPRL2011,GuenotPRA2012}. 
This has stimulated a number of theoretical investigations concerning the role of the long-range Coulomb potential
\cite{KlunderPRL2011,ZhangPRA2010,ZhangPRA2011}, 
many-electron screening effects 
\cite{GuenotPRA2012,KheifetsPRL2010,MoorePRA2011} and 
electron localization 
\cite{ZhangPRA2011_2}. 
The delayed response of atomic systems can be decomposed as the sum of two terms \cite{KlunderPRL2011,NagelePRA2011}:
\begin{align}
\tau_A=\tau_W+\tau_{CC},
\label{delays101}
\end{align}
corresponding to the Wigner-delay of the photoelectron wave packet \cite{WignerPR1955,YakovlevPRL2010} and 
the continuum--continuum delay from the laser-probe process \cite{KlunderPRL2011}. 
Eq.~(\ref{delays101}) is accurate for simple single-electron systems 
\cite{NagelePRA2011,DahlstromCP2012}, 
but its extension to many-electron systems has been predicted to be difficult \cite{IvanovPRL2012}.
It is of great importance for attosecond metrology to understand the origin of these time delays 
\cite{deCarvalhoPR2002,DahlstromJPB2012}, 
but also to be able to account for them in the most accurate way. 
Numerical experiments, based on the propagation of the time-dependent Schr\"odinger equation,  
serve as an important tool to estimate the delays in photoionization, 
but inclusion of all many-electron interactions is not possible at this time 
\cite{KheifetsPRL2010,MoorePRA2011,NagelePRA2012},  
except in the restricted case of helium   
\cite{PazourekPRL2012}. 
In these demanding numerical experiments, 
the fast spatial extension of the photoelectron wave packet  
leads to difficulties for the analysis of the ionization process, 
due to the artificial boundary of the computational box. 
%

\begin{figure}
\begin{center}
	\includegraphics[width=0.65\textwidth]{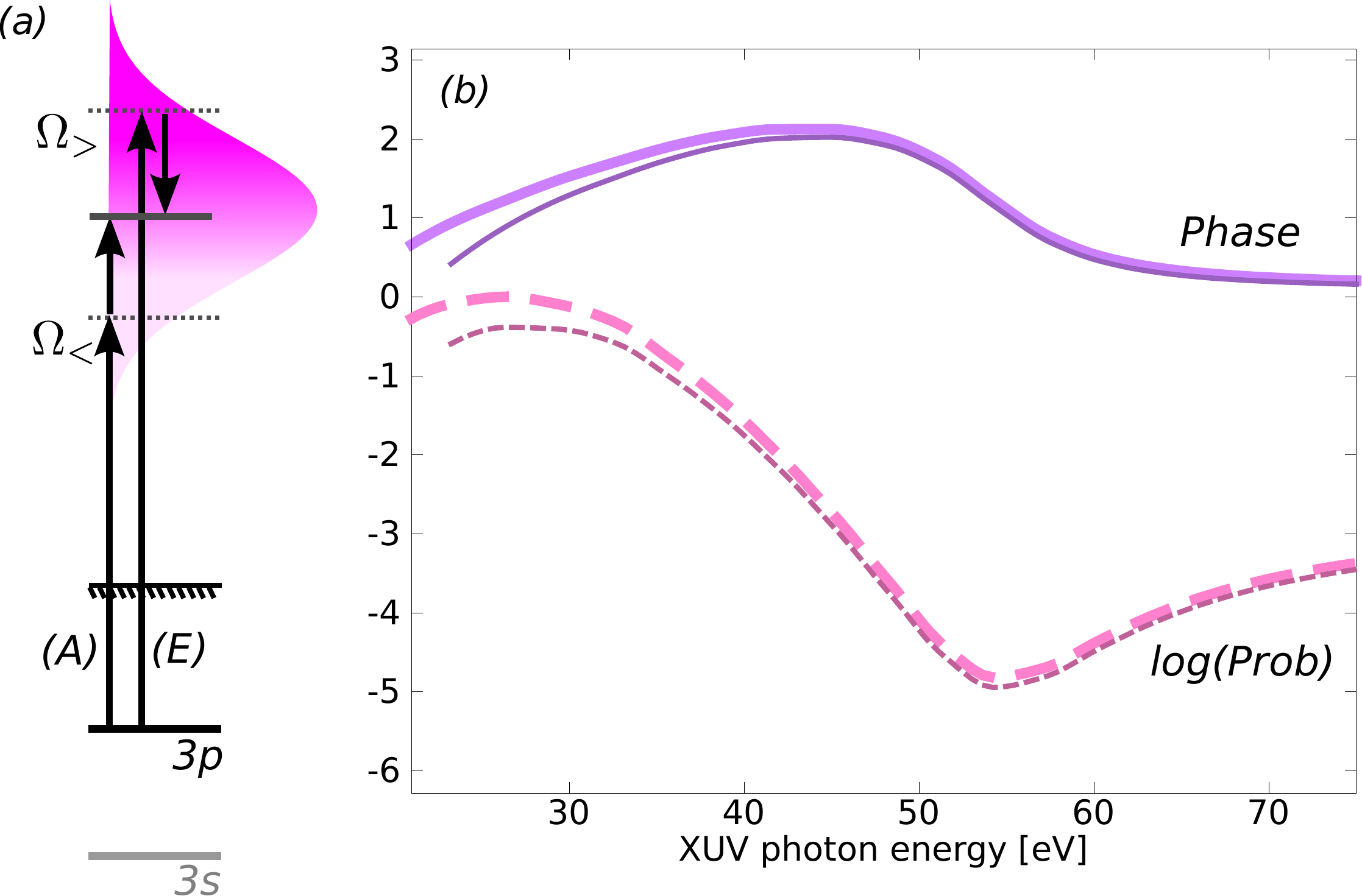} \\
	\caption{
(a) Photon picture of laser-assisted photoionization from the outer-most orbital of argon ($3p$).
(b) Phase (full) and normalized probability (dash) calculated from the two-photon matrix elements. 
The thick curves correspond to absorption of a probe photon (A), 
while the thin curves correspond to emission of a probe photon (E). 
The probe photon is $1.55$~eV.  
} 	
\label{fig1}
\end{center}
\end{figure}

In this Letter we present a method to compute 
complex amplitudes for this class of above-threshold ionization (ATI) transitions in a {\it time-independent} formalism. 
These complex amplitudes are then used to determine the atomic delay in laser-assisted photoionization 
\cite{DahlstromCP2012,DahlstromJPB2012} 
(also referred to as the ``streaking delay'')
for electrons from the outer-most $n$-shell in neon and argon atoms.  
Furthermore, we gauge the validity of Eq.~(\ref{delays101}) 
and discuss the role of multiple ionization channels. 
Correlation effects are accounted for ``all orders'' of single-particle excitations, 
including the non-local exchange interaction and ground-state correlation.
Our analysis is based on the dominant class of two-photon processes, where  
one XUV photon is absorbed from the attosecond pulse and one IR probe photon is exchanged.
The XUV photon frequencies must then differ by two probe photons 
in order to reach the same final state, as can be seen in Fig.~\ref{fig1}~(a) \cite{DahlstromCP2012}. 
The photoelectron is thus probed at different intermediate kinetic energies, 
in close analogy with {\it spectral shear interferometry},
yielding information about the spectral phase-variation of the attosecond pulse \cite{MullerAPB2002}, 
but also including a characteristic response of the electronic wave packet \cite{TomaJPB2002}.   
Interestingly, these two-photon amplitudes depend rather weakly on the probe-step of the process  
in the high XUV energy range, as shown in Fig.~\ref{fig1}~(b) for the case of ionization from the $3p$ orbital in argon. 
The gross features of the two-photon process can, therefore, be identified already in the XUV-photon matrix element, 
{\it e.g.} a ``Cooper minima'' at an XUV energy of $\sim 55$\,eV 
as expected from the $3p$ orbital in argon \cite{MauritssonPRA2005}.
Around this energy, the phase of the matrix element exhibits a non-trivial behavior 
as the dominant ionization rate is shifted from the $3p\rightarrow \epsilon d$  to the $3p\rightarrow \epsilon s$ channel. 
Clearly, this complex region is ideally suited for a quantitative test of Eq.~(\ref{delays101}). 
%


\section*{Method}
\begin{figure}
\begin{center}
	\includegraphics[width=0.65\textwidth]{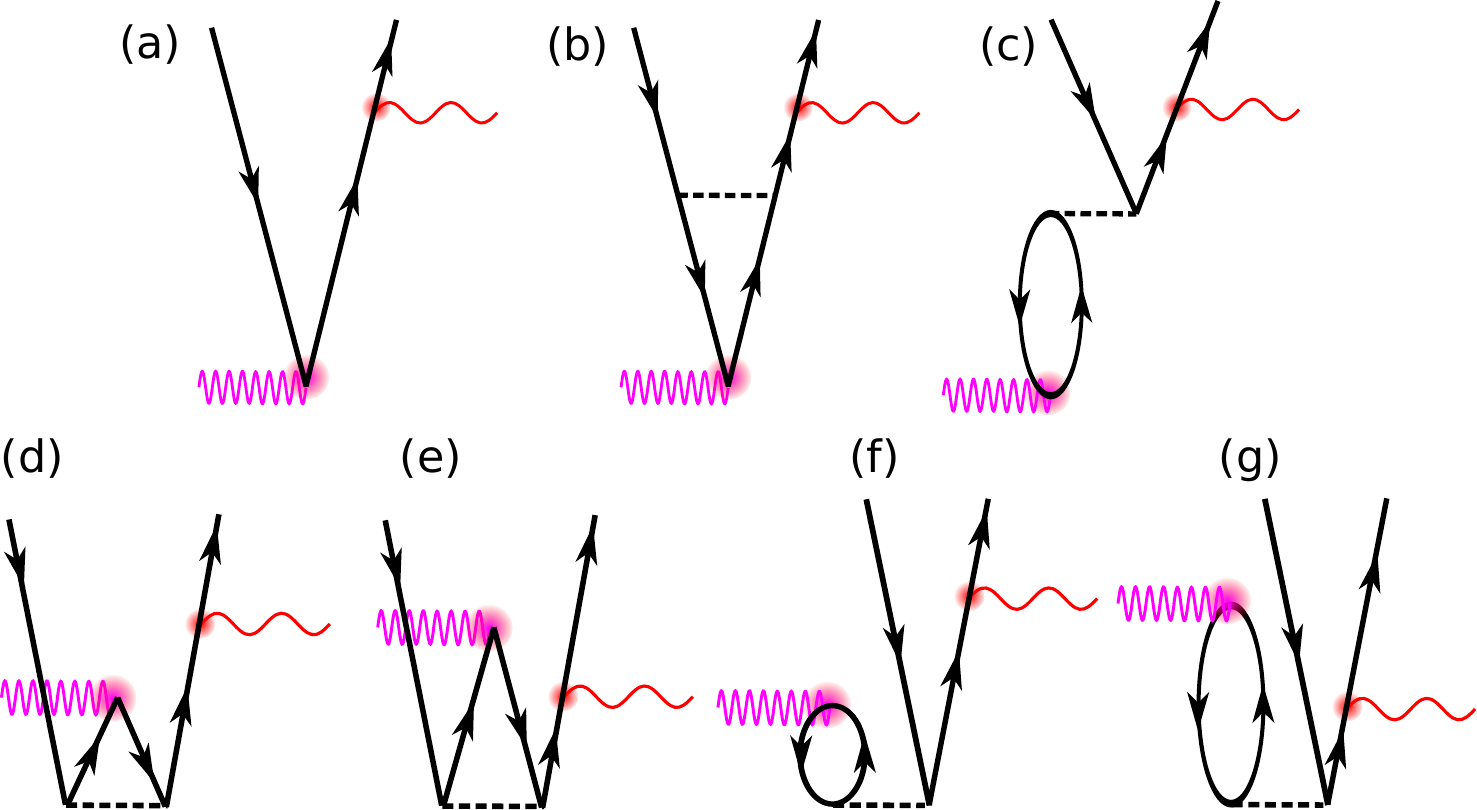} \\
	\caption{
(a) Absorption of XUV (pump) and  IR (probe) photons by the photoelectron.
(b) Direct and (c) exchange interaction between the photoelectron and the remaining electrons in the core. 
(d)-(g) Ground state correlation interactions. 
The XUV photon and the IR photon are indicated by the fast wiggle and slow wiggle respectively, while the interaction between the photoelectron and the remaining hole in the atom is a dashed line. 
}
\label{fig2} 	
\end{center}
\end{figure}

The correlated photoelectron and ion are represented using the perturbation diagrams shown in Fig.~\ref{fig2}, 
including linear screening for the absorption of the XUV photon \cite{HuillierPRA1987}. 
It is convenient to rewrite the two-photon transitions in Fig.~\ref{fig2} using a perturbed wavefunction 
(PWF), $\ket{\rho_{\Omega,a}}$, 
for absorption of one XUV photon, $\Omega$, from an initial atomic orbital, $\ket{a}$, with energy $\epsilon_a$. 
The complex two-photon transition amplitude can then be expressed as a one-photon dipole matrix element 
from this intermediate PWF to the final state, $\ket{s}$, through absorption of one probe photon, $\omega$,
\begin{align}
M_{s,\omega,\Omega,a}=&
\braOket{s}{d_\omega}{\rho_{\Omega,a}},
\label{twophotonPWF}
\end{align}
where global energy conservation for absorption of both photons is imposed: 
$\epsilon_s-\epsilon_a=\Omega+\omega$. 
The photoelectrons are detected along the polarization direction of the fields, 
namely the quantization axis, $\hat z$. 
Note that the dipole matrix element in Eq.~(\ref{twophotonPWF}) is non-trivial because it describes an ATI process, {\it i.e.} a transition between two wavefunctions of continuum character.

\begin{figure}
\begin{center}
	\includegraphics[width=0.65\textwidth]{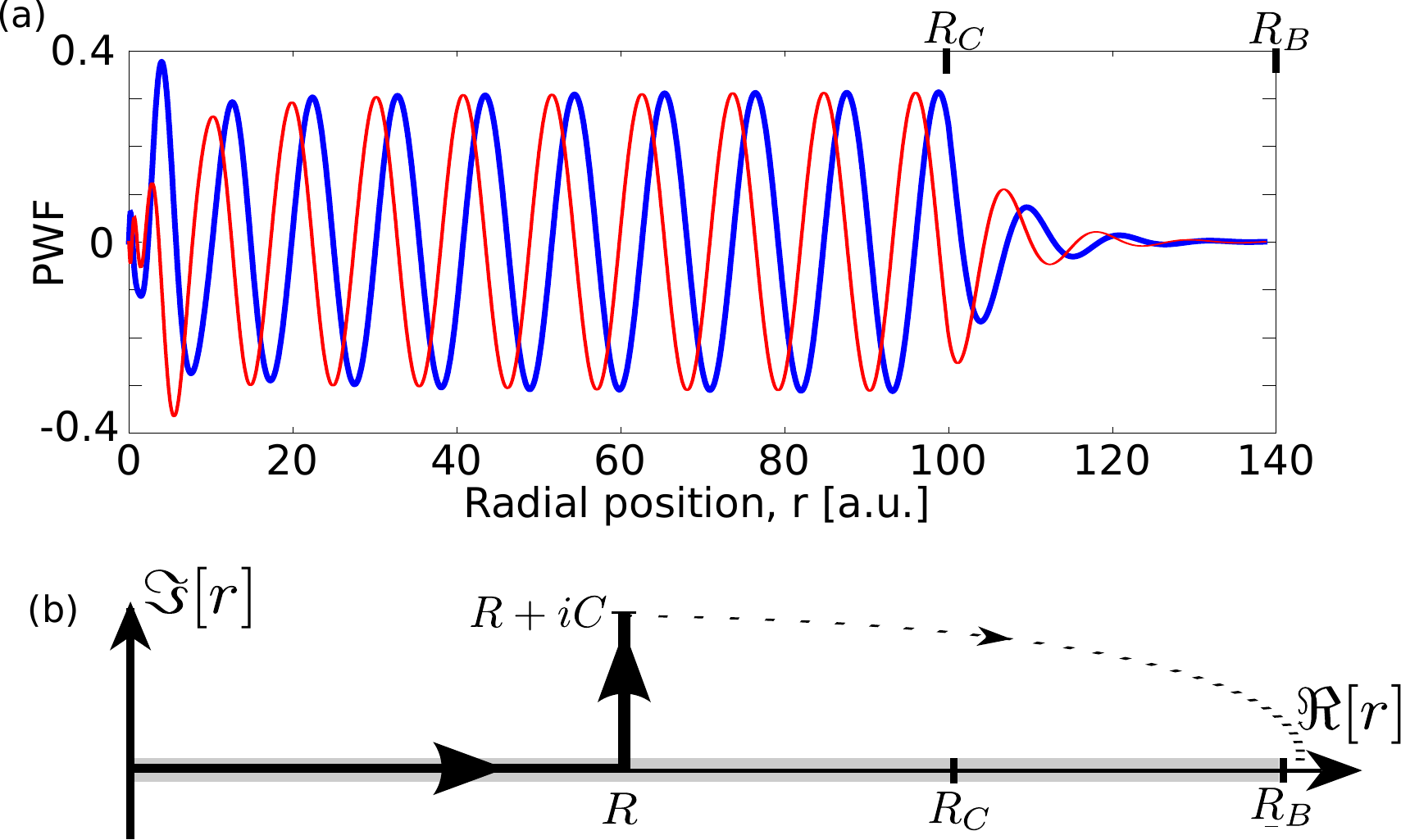} \\
	\caption{
(a) Typical outgoing photoelectron wave packet (PWF), 
following absorption of one XUV photon, with real (red) and blue (imaginary) components shown.
The damping of the PWF for $r>R_C=100$ Bohr radii is due to the exterior complex scaled basis set. 
(b) Radial integration path used to evaluate the dipole transition in Eq.~(\ref{twophotonPWF}). 
Note that $R<R_C$, which implies that the ``break-point'' occurs in the unscaled region of space.
} 	
\label{fig3}
\end{center}
\end{figure}

The PWF is setup in a Hartree-Fock (HF) basis with exterior complex scaling in the radial dimension \cite{SimonPLA1979}:
\begin{align}
r\rightarrow 
\left\{
\begin{array}{lr}
r, 				& 0 < r < R_C	\\
R_C + (r-R_C)e^{i\varphi}, 	& R_C \le r,
\end{array}
\right. 
\label{ECS}
\end{align} 
where the complex scaling starts at $r=R_C$, located far away from the atomic core. 
Typical scaling parameters are: $R_C=100$ Bohr radii and $\varphi=43$ degrees.
This allows for the photoelectron to reach its asymptotic form 
{\it before} entering the exterior complex scaled region, as seen in Fig.~\ref{fig3}~(a). 
A broad range of correlation effects between the photoelectron and the ion can then  
be accounted for using infinite-order, many-body perturbation theory (MBPT). 
%
%
We construct self-consistent equations for the 
forward-propagating ($+$) and backward-propagating ($-$) 
PWF using the diagrams in Fig.~\ref{fig1}: 
\begin{align}
\ket{\rho_{\Omega,a}^\pm} = \ 
\ket{\rho_{\Omega,a}^{(0)\pm}}
-
\sum_t^{exc}\frac{\ket{t}}{(\epsilon_a-\epsilon_t\pm\Omega)} \times
\nonumber \\ 
\sum_b^{core}
\Big\{  \ 
 \braOket{b,t}{r_{12}^{-1}}{a,\rho_{\Omega,b}^\pm}   -  \braOket{b,t}{r_{12}^{-1}}{\rho_{\Omega,b}^\pm,a} 
\nonumber \\ + \ 
      \braOket{\rho_{\Omega,b}^\mp,t}{r_{12}^{-1}}{a,b} -  \braOket{\rho_{\Omega,b}^\mp,t}{r_{12}^{-1}}{b,a} \  \Big\},
\label{dysonlike}
\end{align}  
where 
\begin{align}
\ket{\rho_{\Omega,a}^{(0)\pm}}=\sum_t^{exc}\frac{\ket{t} \braOket{t}{d_\Omega}{a} }{\epsilon_a-\epsilon_t\pm\Omega},
\label{uncorrelatedWF}
\end{align}
is the zeroth-order PWF, {\it i.e.} 
before interaction with the IR photon and with no correlation. 
The sum over excited states ($exc$) in Eqs.~(\ref{dysonlike}-\ref{uncorrelatedWF}) includes bound and continuum states. 
The finite size of the computational box, $R_B=140$ Bohr radii, ensures that the HF basis remains discrete,  
while the use of exterior complex scaling results in complex eigenenergies for the continuum states 
and, therefore, the correct outgoing nature of the photoelectron wave packet (inside the unscaled region). 

Further, in Eq.~(\ref{dysonlike}), 
the terms on the second line correspond to direct and exchange interactions; 
and the terms on the third line represent the ground-state correlation. 
We can solve Eq.~(\ref{dysonlike}) by numerical iteration including all interacting core orbitals, $\ket{b}$, of the atom,
but for the XUV photon energies of interest in this Letter, it is sufficient to include only the outer-most $n$ orbitals. 
Once convergence is found, the forward-propagating PWF is used 
to evaluate the two-photon matrix element in Eq.~(\ref{twophotonPWF}). 
The backward-propagating PWF is only required for a self-consistent treatment 
of ground-state correlation effects in Eq.~(\ref{dysonlike}), c.f. Ref.~\cite{Martensson-PendrippJP1985}.
Also, a PWF for the process where the IR photon is absorbed first, {\it i.e.} time-ordered before absorption of the XUV photon,
can be constructed in a similar fashion, 
but we have verified that these processes are weaker and can be omitted to good approximation. 

%

In order to evaluate Eq.~(\ref{dysonlike}),  
we apply the closed-shell HF approximation. 
Then, we remove the monopole contribution of a vacancy in the atom, say $\ket{a}$, 
from the MBPT operator, $r_{12}^{-1}$, 
to form modified HF equations
that include a ``projected-hole'' potential \cite{LindrothJPB1993}. 
This results in two important consequences: 
(I) the long-range interaction between the photoelectron and the ion, 
due to the monopole term, is included already in the modified HF basis; 
(II) the remaining part of the $r_{12}^{-1}$ interactions are short-ranged. 
This method can be extended also to more complex systems, such as molecules, 
provided that the asymptotic form of the PWF will be determined by this long-range monopole term.    
The final state, $\ket{s}$, is not correlated in our analysis
because there are no interactions (dashed lines) 
after absorption of the probe photon in Fig.~\ref{fig2}.
The exact form of the final state is not critical since    
the absolute phase-shift cancels out when constructing the atomic delay 
\cite{DahlstromCP2012,DahlstromJPB2012}, 
but inter-orbital interactions after the probe step may, none-the-less, 
present an interesting case for further studies.


In the complex scaled region, $R_C<r<R_B$, the photoelectron wave packet is exponentially decaying, 
as observed in Fig.~\ref{fig3}~(a). 
Therefore, it can not be directly used for the transition to the final state in Eq.~(\ref{twophotonPWF}), 
but it is possible to solve this problem using the analytical properties of Coulomb waves.
The ATI matrix element is evaluated by changing 
the path of radial integration in the complex plane {\it before} entering the complex scaled region, as sketched in Fig.~\ref{fig3}~(b). The first part is integrated from $r=0$ to $R$ using a numerical final state and PWF. Both functions are then matched to  Coulomb functions at this ``break-point'' and the next part of the integral is evaluated from $R$ to $R \pm iC$ using these analytical functions. The remaining integral path, sketched as a dashed curve, can be neglected provided that $C$ is large enough, 
typically a few hundred Bohr radii. 
We have computed the integrals for several different break-points in order to verify this procedure. 
\begin{figure}
\begin{center}
	\includegraphics[width=0.65\textwidth]{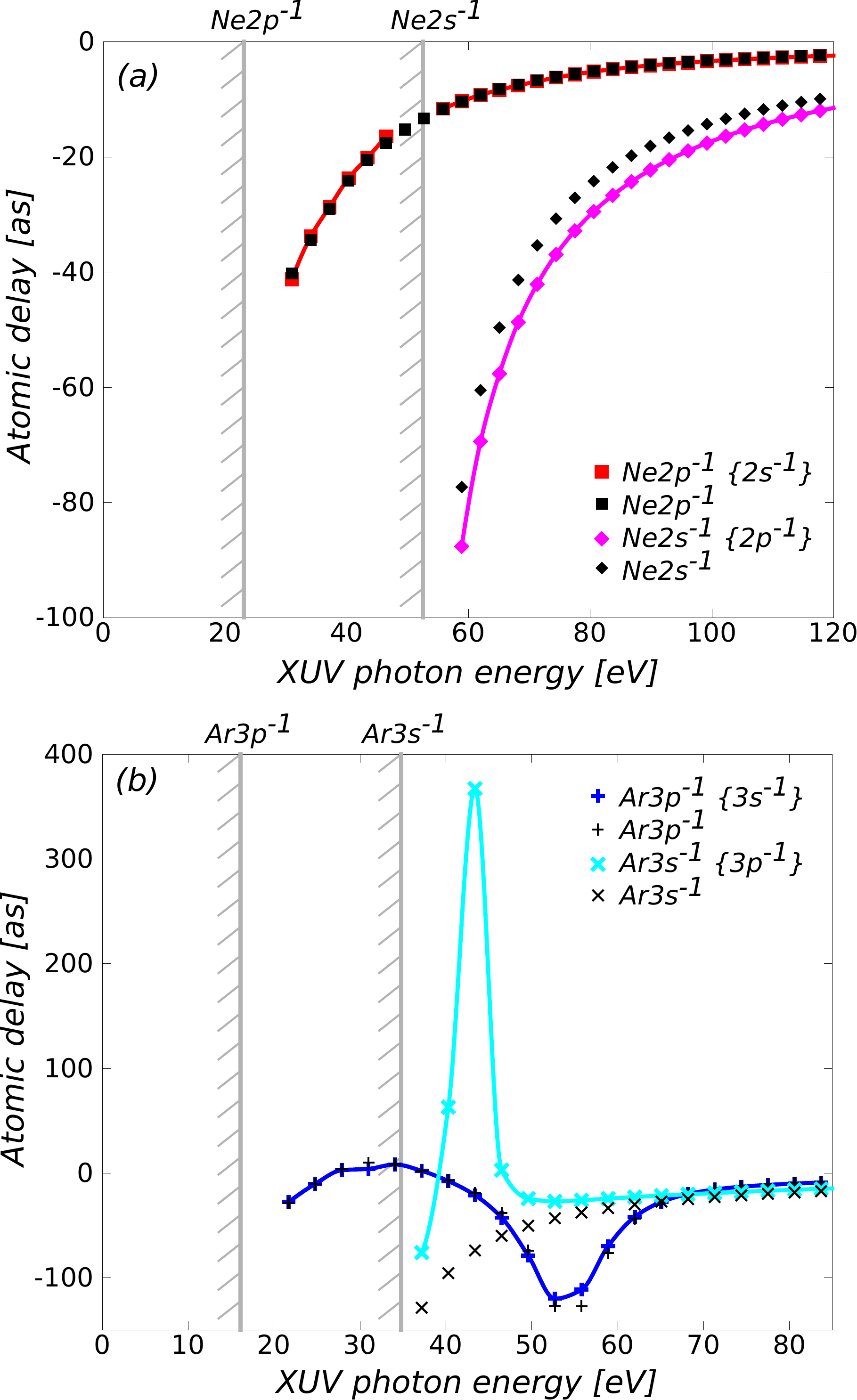} \\
	\caption{
Atomic delays from the outer-most orbitals 
of (a) neon and (b) argon atoms using a probe of 1.55\,eV. 
The data including inter-orbital correlation is outlined, 
while the single-active orbital approximation is marked by black symbols. 
HF ionization thresholds are shown.
} 
\label{fig4}	
\end{center}
\end{figure}
%

\section*{Results}
In Fig.~\ref{fig4}, we present the atomic delays 
for photoelectrons from the outer-most $n$ orbitals of (a) neon and (b) argon atoms.
We find that the atomic delays from the {\it outer} orbitals, $2p$ and $3p$ respectively,
are mostly insensitive to the coupling with the inner orbitals.
This demonstrates the validity of the single-active orbital approximation for outer electrons.  
In contrast, the delays from the {\it inner} orbitals, $2s$ and $3s$ respectively, 
are strongly altered by the coupling to the outer orbitals
and it is required to venture beyond a single-active orbital. 
This inter-orbital correlation leads to an induced delay of a few attoseconds in the entire energy range for neon.
The {\it relative} delay between the two ionization channels is $\sim12$\,as for an XUV photon energy of $\sim105$\,eV, 
which is close to the value of $10.2\pm1.3$\,as obtained using the time-dependent R-matrix approach 
by Moore et al \cite{MoorePRA2011}. 
Intriguingly, these theoretical values are still too small 
to properly explain the $21\pm5$\,as measured experimentally by Schultze et al \cite{SchultzeScience2010}. 
Our approach can be extended to include additional correlation processes \cite{Amusia1990} 
allowing for further investigation of this discrepancy. 

In the case of argon, 
we observe sharp delay structures close to the ``Cooper minima'' from the $3p$ and $3s$ orbitals respectively.
The delay peak from the $3s$ orbital is only observed in the correlated calculation,  
as predicted by Kheifets in Ref.~\cite{GuenotPRA2012}. 
The atomic delay difference between the $3p$ and $3s$ orbitals is $\sim 78$\,as at an XUV photon energy of $\sim37$\,eV,
which is consistent with the experimental value of $100\pm50$ presented in Ref.~\cite{GuenotPRA2012}.

\begin{figure}
\begin{center}
	\includegraphics[width=0.65\textwidth]{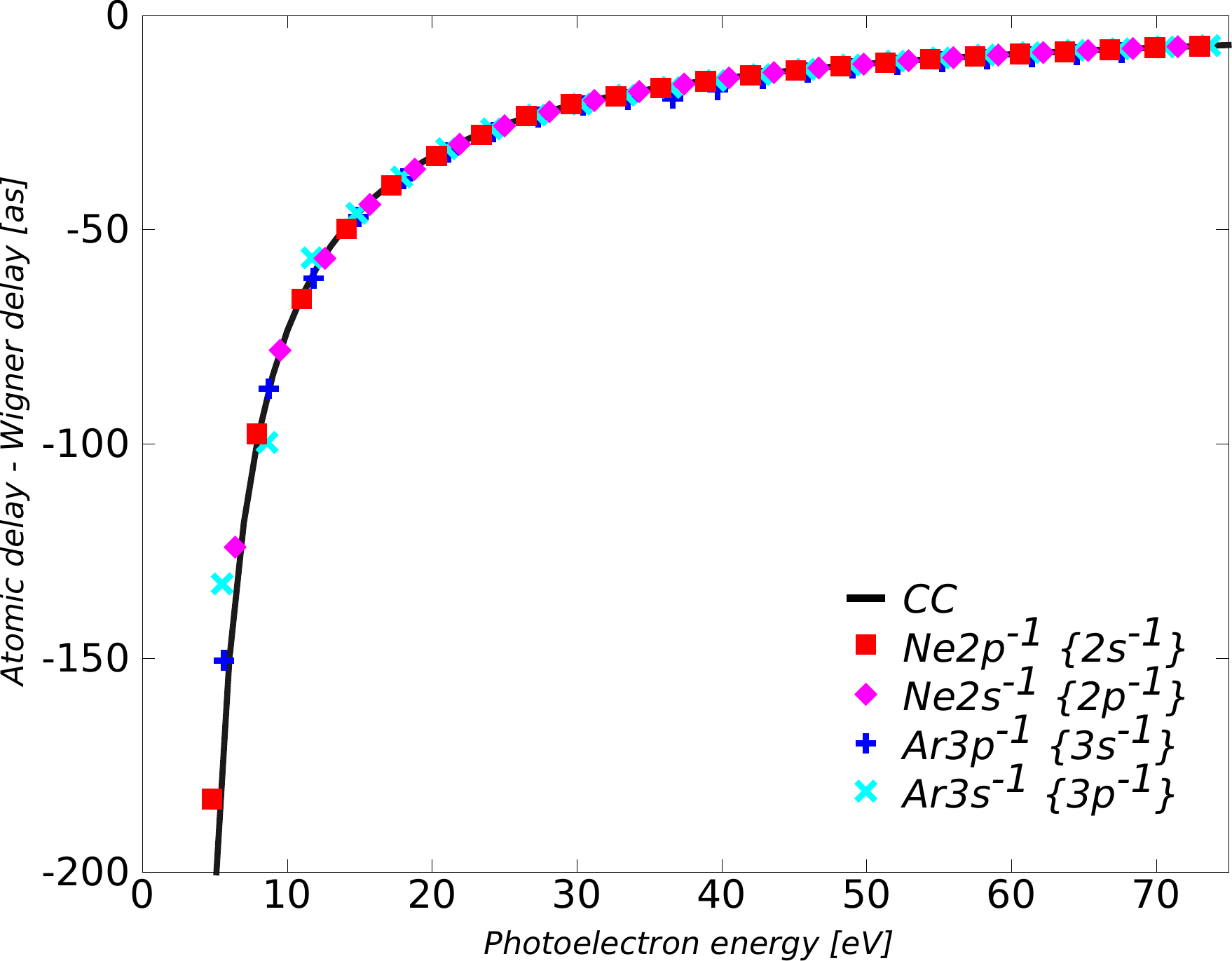} \\
	\caption{
The ``exact'' continuum--continuum delay is determined for all outer-most $n$ shell electrons in neon and argon as 
$\tau_{CC}\equiv\tau_A-\tau_W$ using Eq.~(\ref{delays101}).
The good agreement with the analytical curve (CC) from Ref.~\cite{DahlstromJPB2012} shows that a meaningful separation of the delays can be made for atomic many-electron systems.  
}	
\label{fig5}
\end{center}
\end{figure}

Finally, in order to gauge the validity of Eq.~(\ref{delays101}),  
we extract the Wigner delay, $\tau_W$, from the PWF  
using scattering states for emission along the polarization axis. 
This delay is then subtracted from the total atomic delay 
in order to obtain the ``exact'' continuum--continuum delay, $\tau_{CC}\equiv\tau_A-\tau_W$, for each ionization process,
see Fig.~\ref{fig5}.  
An analytical curve from the asymptotic approximation \cite{DahlstromJPB2012} 
shows good agreement for all ionization processes in neon and argon.
The large delay peaks of several hundred attoseconds in argon 
amount to rather small deviations of $\tau_{CC}$, which are barely visible in Fig.~\ref{fig5}. 


\section*{Conclusions}
In conclusion, 
we have developed a powerful approach to include correlation effects in laser-assisted photoionization. 
The delayed atomic response, $\tau_{A}$, is calculated  
from first principle for all outer-most $n$-shell electrons in neon and argon atoms.  
Future analysis of attosecond pulse structures based on 
RABITT \cite{MullerAPB2002}, PROOF \cite{ChiniOE2010} and FROG-CRAB \cite{MairessePRA2005}
should include $\tau_A$ in order to avoid 
errors of several hundred attoseconds close to atomic features, such as Cooper minima. 
The calculation of the continuum--continuum transition 
relies on the asymptotic form of the intermediate wave packet, but only far away from the atomic core, 
where it is known to be an outgoing Coulomb function. 
In this way, we have demonstrated the validity of Eq.~(\ref{delays101}) also for correlated atomic systems.  
Since the method presented here builds on well-developed many-body methods, 
it can also be extended to other systems where the random phase approximation (RPA) has been applied previously 
and the general form of the asymptotic continuum wave packet is known. 
As an example, 
it would be interesting to study the role of an IR probe field present in photoionization of $N_2$,  
where calculations have been performed within the RPA covering the highest four occupied molecular orbitals 
\cite{PhysRevA.61.032704}.


Financial support from 
the Swedish Research Council 
and
the Wenner-Gren Foundation 
is acknowledged. 


\bibliography{RPLno1}
\bibliographystyle{unsrt}
\end{document}